\documentclass[times,doublespace]{article}
\usepackage{amssymb}
\usepackage[all,knot,arc,import,poly]{xy}
\usepackage{graphicx}
\usepackage{amsmath}
\usepackage{color}
\usepackage{enumerate}
\usepackage{hyperref}
\usepackage{tikz}
\usetikzlibrary{positioning,shapes,arrows, snakes,fit}
\usetikzlibrary{matrix}
\usepackage{graphicx}
\usepackage{subfig}
\usepackage{natbib}

\begin{document}


\textbf{\Large On the combination of omics data for prediction of binary outcomes}

Mar Rodr\'{i}guez-Girondo$^{1*}$, Alexia Kakourou$^{1*}$, Perttu Salo$^{2}$, Markus Perola$^{2}$, Wilma E. Mesker$^{3}$, Rob A. E. M. Tollenaar$^{3}$,  Jeanine Houwing-Duistermaat$^{1,4}$, Bart J.A. Mertens$^{1}$\\
$^{*}$ These authors contributed equally to this work.
\bigskip

\noindent
$^{1}$Department of Medical Statistics and Bioinformatics, Leiden University Medical Center, Leiden, The Netherlands.
$^{2}$National Institute For Health and Welfare, Helsinki, Finland.
$^{3}$Department of Surgery, Leiden University Medical Center, Leiden, The Netherlands.
$^{4}$Department of Statistics, Leeds University, Leeds, United Kingdom.

\begin{abstract}
Enrichment of predictive models with new biomolecular markers is an important task in high-dimensional omic applications. Increasingly,  clinical studies include several sets of such omics  markers available for each patient, measuring different levels of biological variation. As a result, one of the main challenges in predictive research is the integration of different sources of omic biomarkers for the prediction of health traits. We review several approaches for the combination of omic markers in the context of binary outcome prediction, all based on double cross-validation and regularized regression models. We evaluate their performance in terms of calibration and discrimination and we compare their performance with respect to single-omic source predictions. We illustrate the methods through the analysis of two real datasets. On the one hand, we consider the combination of two fractions of proteomic mass spectrometry for the calibration of a diagnostic rule for the detection of early-stage breast cancer. On the other hand, we consider transcriptomics and metabolomics as predictors of obesity using data from the Dietary, Lifestyle, and Genetic determinants of Obesity and Metabolic syndrome (DILGOM) study, a population-based cohort, from Finland.
\end{abstract}

Keywords: \textit{prediction; classification; combination; augmented prediction; double cross-validation; regularized regression}

\section{Introduction\label{section intro}}
Proteomics,  metabolomics and related omics research fields are revolutionizing bio-molecular research by the ability to simultaneously profile many compounds within either patient blood, urine, tissue or other. Increasingly,  clinical studies include several sets of such omics  measures available for each patient, measuring different levels of biology. In the last decade, much work has been done on accommodating one single high-dimensional source of (omic) predictors for prognosis. Nowadays, one of the main challenges in predictive research is the integration of different sources of omic biomarkers for the prediction of health traits.

Prediction with several omic sources involves a number of difficulties. First of all, omic sets of predictors are typically high-dimensional ($n<p$, $n$ sample size and $p$ the number of predictors), and correlation between features is typically high which requires the use of model building techniques beyond classical regression-based methods. Furthermore, several potential predictor sets measured on the same subjects may share (part of) the underlying biological information, which thus introduces correlation between the distinct omic sources. Moreover, differences on scales, normalization procedures and other technical issues inherent to omic research may play a role when trying to integrate information provided by several omic sources. These issues may dramatically affect the gain in predictive ability of a hypothetical `naive' combination, based on stacking all the available features, and ignoring their different origin, specially in high-dimensional settings. For example, one of the sources of predictors may be obscured by another one due to their different dimensionality, noise level and correlation between them.

In this work, we propose to replace the original (high-dimensional) sources of predictors by their corresponding predicted values of the outcome based on single-source prediction models and to combine those, with the intention of outperforming single-omic prediction models. This approach has become relatively popular nowadays for combining predicted values obtained from different methods but based on a common source of predictors \citep{Leblanc, superlearner, Alexia} but it has been less applied in the context of combination of predictions of a common outcome based on different sources of predictors \citep{Mertens2}. An important issue when fitting a model for combining predictions which are themselves fitted is that it requires the calibration of each of the single-omic based predictions as well as the combined model using the same set of observations. In this situation, the calibration of the resulting combination will depend on the prior calibration using the single sources. This issue can be handled by appropriate use of cross-validation techniques.

The former setting refers to the parallel combination of omic sources, in the sense that both single-omic predictions are jointly considered, without imposing any `a priori' hierarchy or different importance among them. Alternatively, combination can be regarded as the sequential augmentation with new biomarkers of previously calibrated prediction models based on a given omic dataset. In this way, the first source of predictors is prioritized. Several examples may motivate such an asymetric approach to combination, such as the addition of more expensive predictors to potentially improve the performance of a prediction rule based on more economic sources or the addition of less reliable sources to models based on stable and well established markers. It seems clear that when a new molecular marker set emerges due to (technical) advances in the field, it must then prove its worth in the face of existing knowledge on the predictive capacity of an established biomarker set. However, if a sequential augmentation outperforms the parallel combination in terms of predictive performance is unclear.

Next, we present and discuss different strategies for the combination and augmentation of single-omic prediction models in the context of classification problems (binary outcome). We discuss their relations and differences and we compare the considered methods by means of the analysis of two real datasets. On the one hand, we revisit the problem of calibrating an early diagnosis tool for breast cancer based on MS-based proteomics profiling. Specifically, we consider the combination of two different sets of predictors, each obtained by different techniques of fractionation of the same spectromety data \citep[see][for details]{Noo,Mertens2}. On the other hand, we consider data from a population-based cohort, the Dietary, Lifestyle, and Genetic determinants of Obesity and Metabolic syndrome (DILGOM) study, sampled from the Helsinki area, in Finland \citep{Inouye1}. In this case, we consider the combination of transcriptomics and metabolomics sources in the prediction of obesity.

\section{Methods}
\subsection{Double cross-validation prediction}

\indent Let the observed data be given by $(\mathbf{y},\mathbf{X}_{1},\mathbf{X}_{2})$, where $\mathbf{y}=(y_{1},\ldots,y_{n})^{\intercal}$ is a binary outcome, $y_{i} \in \{0,1\}$ for $i=1,\ldots,n$ independent individuals and $\mathbf{X}_{1}$ and $\mathbf{X}_{2}$ are two matrices of dimension $n\times p$ and  $n \times q$, respectively, representing two omic predictor sources with $p$ and $q$ features. We assume that we are in a high-dimensional setting ($p,q>n$) and that our objective is to enrich single omic predictions $p_{ik}=\widehat{P}(y_{i}=1|\mathbf{X}_{ki})$, $k=1,2$ by the combined use of the two distinct sources.

Two crucial difficulties in high-dimensional prediction problems are the control of the optimal level of shrinkage (or in general, any tuning parameter $\lambda$ associated with the statistical model $f$ used to obtain estimates of $\mathbf{y}$ based on a single high-dimensional source of predictors) and the quantification of the error of the resulting predictions in new data. Double cross-validation algorithms \citep{Stone, Breiman, Jonathan, Mertens0, Mertens1, Mertens2}, consisting of two (or more) nested loops, allow to handle both issues. In the inner loop a cross-validated grid-selection is used to determine the optimal prediction rule, i.e., for model selection, while the outer loop  is used to estimate the prediction performance by application of models developed in the inner loop part of the data (training sets) to the remaining unused data (validation sets). In this manner, double cross-validation is capable of jointly calibrating and assessing models in a predictive sense,  while also avoiding the bias in estimates of predictive ability which would result from use of a single-cross-validatory approach only.

Next, we present two approaches for combination of omic-based predictions of binary health traits, both based on the replacement of the original (high-dimensional) sources of predictors by their corresponding estimated values of the outcome based on single-source prediction models and the double cross-validation principle. On the one hand, we consider {\it parallel} combination methods, in which the outer cross-validation loop of the `double' cross-validation procedure is used to calculate, in a first step, predictions of the outcome $\mathbf{y}$ of interest based on each of the omic sources of predictors, $\mathbf{X}_{1}$ and $\mathbf{X}_{2}$, which are combined in a second step. On the other hand, we consider a {\it sequential} combination of $\mathbf{X}_{1}$ and $\mathbf{X}_{2}$ in which the double cross-validated predictions of $\mathbf{y}$ based on $\mathbf{X}_{1}$ (considered as primary source) enter as an offset term (not refitted) in a second step in which $\mathbf{X}_{2}$ is evaluated as an additional set of variables to predict $\mathbf{y}$.

\subsection{Parallel combination of predictions}
\indent Firstly, we consider a parallel combination approach based on replacing the original sets of predictors $\mathbf{X}_1$ and $\mathbf{X}_2$ with the sets of their corresponding estimated class probabilities $\mathbf{p}_1=(p_{11},...,p_{1n})^{\intercal}$ and $\mathbf{p}_2=(p_{21},...,p_{2n})^{\intercal}$. In a first stage, the double cross-validated probabilities $\mathbf{p}_1$ and $\mathbf{p}_2$ are estimated by calibrating the prediction model using each single predictor source $\mathbf{X}_1$ and $\mathbf{X}_2$ as input variables. Subsequently, we combine the double cross-validated class probabilities either by considering convex combinations of the estimated class probabilities or by using them as new input variables for the construction a final combined model.

\subsubsection{Convex combination via linear mixtures}

\indent A simple way to combine the cross-validated estimates $\mathbf{p}_1$ and $\mathbf{p}_2$ is to consider linear mixtures of the separately calibrated class probabilities
\begin{equation}
\mathbf{p}_{C} = w \mathbf{p}_{1} + (1-w) \mathbf{p}_{2}
\end{equation}
with $\mathbf{p}_C = (p_{C1},...,p_{Cn})^{\intercal}$ the newly combined class probabilities vector and $w$ some number in the interval $[0,1]$. Since different choices of $w$ result in different prediction rules, we should choose $w$ so that it optimizes the final prediction rule. The parameter $w$ can be considered as the optimal balance between the predictions based on the two distinct sources $\mathbf{X}_1$ and $\mathbf{X}_2$. Note that the two extreme choices, for which $w$ is either 0 or 1, result in excluding $\mathbf{X}_1$ or $\mathbf{X}_2$ completely from estimating the combined class probabilities vector $\mathbf{p}_{C}$.

In many applications, it has been observed that the predictive performance of linear combinations of predictors are often insensitive to the exact values of their weights $w$, i.e., quite large deviations from the optimal set of weights $w$ often lead to predictive performance not substantially worse than those obtained using optimal weights. This phenomenon has been termed as the flat maximum effect \citep{Hand1, Hand2}. Specifically, if the correlations between $\mathbf{p}_{1}$ and $\mathbf{p}_{2}$ are high, the simple average ($w=0.5$) will be highly correlated with any other weighted sum of $\mathbf{p}_{1}$ and $\mathbf{p}_{2}$, and hence the choice of weights will make little difference.Thus, averaging across the estimated class probabilities is expected to result in improved estimates in many omic settings. The main advantage of this approach lies in its simplicity, due to the fact that no further optimization or cross-validatory scheme is required in order to obtain an unbiased estimate of the predictive performance, since the double cross-validatory nature of the estimated class probabilities is preserved entirely.

\subsubsection{Model-based combination}

\indent A somewhat more sophisticated way to combine the cross-validated estimates, in a parallel fashion, is to fit a (semi)parametric model, such as an ordinary logistic regression model, to the set of double cross-validated class probabilities $(\mathbf{p}^{1},\mathbf{p}^{2})$. In this case, the original set of predictors is replaced with the set of estimated class probabilities, reducing the dimensionality of the original data to a low-dimensional space. The final combined class probabilities can then be derived by fitting the logistic model

\begin{equation}
\text{logit}(p_{Ci})=\alpha+\beta_{1}\text{logit}(p_{1i})+\beta_{2} \text{logit}(p_{2i})
\end{equation}
where $\text{logit}(p_{ki})=\log(\frac{p_{ki}}{1-p_{ki}})$. To maintain the double cross-validatory nature of the predictive performance evaluation we embed the logistic model calibrations within an additional single cross-validatory loop, leaving out each cross-validated pair ($p_{1i}$,$p_{2i}$) in turn and fitting the logistic model using the remaining pairs to obtain the final (cross-validated) class probabilities $\mathbf{p}_{C}=(p_{C1},\ldots,p_{Cn})^{\intercal}$.

Attention must be brought at this point to the fact that, in the case of model-based combination, the class probabilities $(\mathbf{p}^{1},\mathbf{p}^{2})$ are not only combined, as in the case of convex combination, but also re-calibrated as suggested by \cite{Cox}. This re-calibration aspect should be taken into account when interpreting the results from this combination approach and when comparing them directly to the performance measures of the calibrated, yet not re-calibrated model, based on $\mathbf{X}_{1}$ or $\mathbf{X}_2$ only. This is of particular importance since fitting a model based on the set of the calibrated estimates from a single predictor source, instead of the single predictor source itself, could alter the final estimates. A more fair comparison thus would be the comparison between the cross-validated predictions of the logistic model combination and the cross-validated predictions of the re-calibrated logistic models using the estimates from the first source only ($\mathbf{p}_1$) or the estimates from the second source only ($\mathbf{p}_2$) such that

\begin{equation}
\text{logit}(p^R_{ki})=\alpha+\beta_{1}\text{logit}(p_{ki})
\end{equation}

for $k=1,2$, with $p_{ki}^R$ the re-calibrated probabilities for the $k^{th}$ source. This type of comparison would assess the extent of the re-calibration effect and would give us insight in whether the improvement in prediction performance is due to combining the cross-validated estimates using each different source or due to re-calibration.

\subsection{Sequential combination of prediction}
An alternative approach to the aforementioned parallel combination approach is to consider the problem of combination of omic predictors in an `asymmetric' manner by sequentially fitting prediction models based on different omic datasets. If we focus on the study of two omic datasets, we can proceed as follows. Firstly, the  double cross-validated predictions of the outcome $\mathbf{y}$ based on the primary source, $\mathbf{X}_{1}$, and a given model specification, $\mathbf{p}_{1}=(p_{11},\ldots,p_{1n})^{\intercal}$ are estimated. Then, in a second step we construct a second  model based on $\mathbf{X}_2$ as predictor and devoted to predict the variation of $\mathbf{y}$ which remains unexplained by $\mathbf{X}_{1}$. In the logistic regression context, this can be implemented by considering $\text{logit}(\mathbf{p}_{1})$ as an offset term in a regularized regression based on $\mathbf{X}_{2}$ for predicting $\mathbf{y}$ as follows:

\begin{equation}
\text{logit}(p_{Ci})=\text{logit}(p_{1i})+f(\mathbf{X}_{2i})
\end{equation}

The first source of omic predictors is hence prioritized by its inclusion as offset, and hence not refitted in the second stage. This means that the second source of predictors $\mathbf{X}_{2}$ is `added' to the previous fit based on $\mathbf{X}_{1}$, since the single-omic prediction $\mathbf{p}_{1}$ is fixed in the second step.

As was the case in the model-based parallel combination presented in Section 2.2.2., the model in expression (2) contains $\mathbf{p}_{1}$ which is fitted itself. Hence, following the lines of Section 2.1., we embed the estimation of $\mathbf{p}_{1}$ in the double cross-validation loop to guarantee a realistic estimation of the predictive performance of the sequential combination. Namely, we leave out each cross-validated $p_{1i}$ and corresponding $y_i$ and $\mathbf{X}_{2i}$ in turn and fit the regularized regression model based on  the remaining observations in $\mathbf{X}_{2}$ to get the final $\mathbf{p}_{C}=(p_{C1},\ldots,p_{Cn})^{\intercal}$.

\section{Performance evaluation}
The performance of prediction models can be summarized in several ways. Traditionally, two aspects have been considered as crucial when evaluating prediction models for binary outcomes: calibration and discrimination \citep[see][for a review]{Steyerberg1}. Calibration refers to the quantification of how close predictions are to the actual outcome, while discrimination focuses on determining to what extent individuals with positive outcome have higher risk predictions than those with negative outcome. The relation and differences among them has been object of extensive research in the past years \citep[see][and references therein]{Pepe}. Beyond calibration and discrimination, other measures such as reclassification \citep{Pencina}, clinical usefulness \citep{Vickers}, have been proposed, but an exhaustive comparison of performance measures falls beyond the scope of this work.

\subsection{Calibration measures}
To evaluate the predictive performance of the different combination strategies described in Section 2, and to compare them with single-omic predictive models, we considered several calibration measures.

Denote by $PRESS(\mathbf{y},\mathbf{p})=\sum\nolimits_{i=1}^n (y_i-p_{i})^2$ the prediction sum of squares based on some arbitrary vector of predictions $\mathbf{p} = (p_{1},...,p_{n})^{\intercal}$. The prediction sum of squares is also denoted as Brier score, and it is usually used for reporting model performance.

Consider $\mathbf{p}_{0}$, the simplest cross-validated predictor of $\mathbf{y}$ based on an intercept-only logistic model, corresponding to the classification rule based on assigning all the observations to the majority class. Denote by $CVSS(\mathbf{p}_{1},\mathbf{p}_{2})=\sum_{i=1}^{n}(p_{1i}-p_{2i})^{2}$ the cross-validated sum of squared differences of two  vectors of predictions $\mathbf{p}_{1}$ and $\mathbf{p}_{2}$.

For any vector $\mathbf{p}$ of predicted values of the outcome of interest, we define:

\begin{equation}
Q^{2}_{\mathbf{p}}=\frac{CVSS(\mathbf{p},\mathbf{p_{0}})}{PRESS(\mathbf{y},\mathbf{p_{0}})}=\frac{\sum_{j=1}^{J}\sum_{i\in S^{(j)}}\left(p_{i}-p_{0i}\right)^2}{\sum_{j=1}^{J}\sum_{i\in S^{(j)}}\left(y_{i}-p_{0i}\right)^2}.
\end{equation}

where the computations of $p_{0i}$, $p_{1i}$ and $p_{2i}$, $i \in S^{(j)}$ for each of the $j = 1,...,J$ random splits $S^{(j)}$ of the sample $S$ is based on the observations not belonging to $S^{(j)}$. Intuitively, $Q^2_{\mathbf{p}_1}$ and $Q^2_{\mathbf{p}_2}$ can be regarded as cross-validation equivalents of the $R^{2}$, in which the performance of the model-based predictions are compared to the naive predictions based on the prevalence of the outcome variable $y$. Analogously, $Q^2_{\mathbf{p}_C}$ represents the predictive ability of the combination of $\mathbf{X}_1$ and $\mathbf{X}_2$, obtained with any of the methods presented in Section 2.

Additionally, we evaluate the calibration of each of the combinations strategies and for each individual source of predictors in terms of the deviance given by

\begin{equation}
Deviance_{\mathbf{p}}= -2\sum\limits_{i=1}^n y_i\log{p_{i}}+(1-y_i)(\log{(1-p_{i}}))=-2\sum\limits_{i=1}^n \log(1-|y_i-p_{i}|)
\end{equation}

and which is evaluated in the cross-validated predicted probabilities.

\subsection{Discrimination measures}
To quantify the discrimination ability of the different methods introduced in Section 2, we use the c-statistic, the most commonly used summary of discrimination. The c-statistic accounts for the proportion of individuals with $y=1$ with higher predicted probability than individuals with $y=0$, among all possible pairs. It also can be defined as the area under the receiver operating characteristic (ROC). Specifically:

\begin{equation}
\text{c-index}_{\mathbf{p}}=\frac{1}{n_{1}n_{2}}\sum\limits_{i\in G_{1}}\sum\limits_{j\in G_{2}}\left[I\left(p_{i}>p_{j}\right)+0.5\times I\left(p_{i}=p_{j}\right)\right]
\end{equation}

where $G_{1}$ and $G_{2}$ are the index sets for $y=1$ and $y=0$, respectively and $n_{1}$ and $n_{2}$ their respective sizes. The c-statistic is a measure of discrimination, that is, of the extent the double cross-validated predictions are higher for individuals in the groups defined by the outcome variable $y$ and it varies from 0 and 1. A value of 0.5 indicates that probabilities are distributed randomly among the two groups given by $y=1$ and $y=0$, while a value equal to 1 indicates that all predicted probabilities for individuals with $y=1$ are higher than the probabilities assigned to individuals with $y=0$. A c-index below 0.5 indicates reverse ordering, i.e., probabilities for individuals with $y=1$ are lower than estimated probabilities for individuals with $y=0$. Because the c-index is invariant under monotone transformations, it provides no information about calibration, however separation among classes defined by the outcome variable $y$ is of great interest in diagnosis applications, for example.

\section{Application}
\subsection{Data presentation}
\subsubsection{Breast cancer data}
\indent The first study we consider is a clinical proteomics study conducted in the Leiden University Medical Center, The Netherlands, which comprises 307 women, from which 105 are breast cancer patients and  202 are healthy controls. In order to classify participants as cancer cases or healthy controls, we use two different subsets of proteins. Both were processed by MALDI-TOF mass-spectrometry but they differ in the techniques used for extraction, which yields different subsets of proteins suitable for detection. On the one hand, we consider 48 protein measures resulting from the use of weak-cation exchange magnetic beads for protein extraction ($WCX$). On the other hand, the use of reversed-phased C18 magnetic beads ($C18$) resulted in 42 different measures \citep[see][for details]{Noo,Mertens2}.

We carried out two distinct analyses using the combination approaches described in Section 2. As a first analysis, we adopted a parallel combination approach, and we derived combinations of $WCX$ and $C18$ bead processing measures based on a linear mixture and model-based combinations, presented in Subsection 2.2.  Additionally, we also constructed a naive combination by stacking all the 90 available features without distinguishing if they come from the same pre-processing method. Secondly, we analyzed the data using the sequential approach revisited in Subsection 2.3. We considered the $WCX$ method as state-of-art technique which is treated  as primary omic source and we evaluated the added value of the features obtained with the $C18$ processing method. Alternatively, we turned around the roles of the available sources by firstly fitting a model based on the $C18$ source of proteomic predictors.

\subsubsection{DILGOM data}
In a second case study, we consider data from a population-based cohort, the Dietary, Lifestyle, and Genetic determinants of Obesity and Metabolic syndrome (DILGOM) study, sampled from the Helsinki area, in Finland \citep{Inouye2,Inouye1}. We are interested in getting insight in the role of serum NMR metabolites measures and gene expression levels in the prediction of obesity (defined in terms of the Body Mass Index (BMI), specifically, an individual was considered to be obese if $BMI\geq30$). The metabolomic predictor data consists of quantitative information on 139 metabolic measures, mainly composed of measures on different lipid subclasses, but also amino acids, and creatine \citep[see][for details]{Inouye1}. The gene expression profiles were derived from Illumina 610-Quad SNParrays (Illumina Inc., San Diego, CA, USA). Initially, 35,419 expression probes were available after quality filtering \citep[see][for pre-processing details]{Inouye2}. In addition to the pre-processing steps described by \cite{Inouye2}, we conducted a prior filtering approach and removed from our analyses those probes with extremely low variation \citep[see][for details on the conducted pre-processing]{Liu}. As a result, we retained measures from 7380 beads for our analyses. The analyzed sample contained $n=406$ individuals for which both types of omic measurements and the outcome of interest were available. From them 78 (19 \%) were obese.

As for the breast cancer data, we performed both parallel and sequential combinations of metabolome and gene expression for the prediction of obesity. Specifically, we firstly obtained a naive, linear mixture, and model-based parallel combinations of transcriptomics and metabolomics. Secondly, we analyzed the data using the sequential approach revisited in Subsection 2.3. We considered the metabolic profile as primary omic source for the prediction of obesity and evaluated the added value of the blood gene expression profiles. Alternatively, we turned around the roles of the omic sources,  first fitting a model based on gene expression and then adding the metabolome.

\subsection{Model choice: Logistic regularized regression}

Several statistical methods are available to derive prediction models of binary outcomes in high-dimensional settings. In this work, we focus on regularized  logistic regression models \citep{Hoerl, leCessie,Tibshirani1, Hastie1}. For a given omic source of predictors $\mathbf{X}$, $f(\mathbf{X})=\mathbf{X}\boldsymbol\beta$ and $\text{logit}(p_{i})=\mathbf{X}_{i}\boldsymbol\beta$, with $p_{i}=\widehat{P}(y_{i}=1|\mathbf{X}_{i})$.
The estimation of $\boldsymbol{\beta}$ is conducted by maximizing the penalized log-likelihood $\sum_{i=1}^{n}\left[y_{i}\log(p_{i})+(1-y_{i})\log(1-p_{i})\right]-\lambda pen(\boldsymbol \beta)$. The penalty parameter $\lambda$ regularizes the $\boldsymbol \beta$ coefficients, by shrinking large coefficients in order to control the bias-variance trade-off. We consider two different (and widely used) penalization types. On the one hand, we use the ridge penalty \citep{Hoerl, leCessie}, with $pen(\boldsymbol \beta)=||\boldsymbol \beta||_{2}^{2}=\sum_{j=1}^{p}\beta_{j}^{2}$, i.e., a \ $\ell_{2}$ type penalty. On the other hand, we consider lasso regression \citep{Tibshirani1}, with $pen(\boldsymbol \beta)=||\boldsymbol \beta||_{1}=\sum_{j=1}^{p}|\beta_{j}|$, i.e., lasso uses a \ $\ell_{1}$ type penalty.

Note that other  model building strategies for prediction of binary outcomes could have been used in this framework, such as the elastic net penalization \citep{Zou} by setting $\alpha=0.5$, boosting methods \citep{Tutz, Buhlmann, Kneib}, or random forests \citep{Breiman} among others.

\subsection{Results}
The main findings from the analysis of the breast cancer data are summarized in Table 1 and Table 2, while Table 3 and Table 4 show the results corresponding to the DILGOM data. The top parts of the tables refer to the results obtained by using ridge regression as model to derive the double cross-validated predictions and the bottom parts of the tables refer to lasso regression. For both datasets, we provide results of the individual performance (in terms of calibration and discrimination) of each of the two considered omic sources, jointly with the evaluation of the previously introduced  strategies for (parallel and sequential) combination. In Tables 1 and 3, the single omic-sources are evaluated in terms of the (non re-calibrated) predictions $\mathbf{p}_{1}$ and $\mathbf{p}_{2}$. Tables 2 and 4 contain the results for each single omic source based on re-calibrated probabilities, along the lines of expression (3).

\subsubsection{Breast cancer data}
In the breast cancer setting, we observe a slightly better performance of the protein fractionation $WCX$ than the alternative $C18$, according to the two studied model specifications (ridge and lasso regression) and regarding both re-calibrated and non re-calibrated results. Both sets of markers show similar and very good performance in terms of discrimination ($\text{c-index}$ around 0.91 for $WCX$ and slightly inferior for $C18$, with $\text{c-index}$ around 0.88). $WCX$ also outperforms $C18$ in terms of Brier score, deviance and $Q^{2}$. In terms of model specification, lasso regression seems to provide better results in the individual evaluation of each of the omic sources (Table 1), specially in terms of calibration, but the two model specifications behave similar when compared in terms of re-calibrated probabilities (Table 2). Interestingly, re-calibration of the single-omic predictions is beneficial when using ridge regression in terms of calibration, while discrimination becomes slightly worse. On the other hand, re-calibration of single-source lasso-based predictions is not beneficial (all the summary measures worsen with re-calibration).

With regard to the combination strategies, we observe that the na\"{i}ve combination (stacking the two sources of predictors and conduct a new regularized regression with common penalty) provides, in general, worse results compared to the alternative strategies for  parallel combination (averaging and model-based), in terms of both, calibration and discrimination measurements. In fact, the na\"{i}ve combination provides worse results than using $WCX$ only.

The model-based parallel combination outperforms the (simpler) parallel combination based on averaging in terms of calibration (smaller values of Brier score and deviance and larger values of $Q_{\mathbf{p}_{C}}^{2}$), for both ridge and lasso specifications. The difference between average and model-based is almost two-fold for ridge regression, while the difference is smaller for lasso regression. ($Q_{C}^{2}=0.286$ for averaging, $Q_{C}^{2}=0.559$ for the logistic regression combination). Alternatively, if we focus on the differences in discrimination ability, averaging seems to show a slightly better performance than the model-based combination. The right part of Table 1 provides the results from a sequential approach to the combination of omic predictors. Firstly, we observe that, especially for the lasso specification, the order of introduction of the predictors influences the resulting performance of the sequential combination. We denote by  $C18|WCX$ the sequential combinations of $WCX$ and $C18$, introduced in Section 2.3., which treats $WCX$ as primary source. Alternatively, $WCX|C18$ refers to the sequential combinations of $WCX$ and $C18$ which considers $C18$ as the primary source. We observe that for the lasso specification, the sequential combination $C18|WCX$ provides better results than $WCX|C18$, both in terms of calibration and discrimination. These results agree with the intuition that the preferable sequential procedure is the one which starts using the omic source which optimizes the performance when considered individually. Results are less conclusive when using ridge regression as model to sequentially generate the predictions. The results are less influenced by the order in the sequential procedure, and the summary measures disagree with regard to the preferable strategy. $WCX|C18$ presents better results in terms of Brier score, deviance and $\text{c-index}$, while the $Q^{2}$ is larger for $C18|WCX$.

Interestingly, the sequential combinations may outperform the parallel model-based alternative. For ridge regression, we observe that the lowest Brier score and deviance are reached by the sequential combination $WCX|C18$ ($BS=0.088$, $Deviance=188.50$ for $WCX|C18$ versus $BS=0.093$, $Deviance=201.00$ for the model-based parallel combination), and also the largest c-index ($\text{c-index}=0.933$ for $WCX|C18$ versus $\text{c-index}=0.922$ for the parallel combination). The model-based parallel combination presents the best performance in terms of $Q^{2}$ ($Q^{2}=0.559$ for the parallel combination versus $Q^{2}=0.537$ for the sequential combination $C18|WCX$). When using lasso regression, the parallel combination outperforms the sequential approach in terms of Brier score, and deviance, while the sequential combination $C18|WCX$ presents the best results in terms of $Q^{2}$ ($Q^{2}=0.703$ for $C18|WCX$ versus  $Q^{2}=0.609$ for the model-based parallel combination). The results in terms of discrimination measured through the c-index are the same for both model-based parallel and sequential $C18|WCX$.

Finally, note that the model-based combinations, both parallel and sequential outperform the single-omic models, even after accounting for re-calibration of those. The average-based parallel combination seems also beneficial when focusing on discrimination, but its appropriateness is questionable when focusing on calibration, specially in terms of $Q^2$, its performance is worse than the observed for single-omic models.

\begin{table}[htbp]
  \centering
  \caption{Breast cancer data}
    \begin{tabular}{cccccccccc}
    \hline
          &       & \multicolumn{2}{c}{Single source} && \multicolumn{5}{c}{Combination methods} \\
          \hline
  &&C18&WCX&&Naive&\multicolumn{2}{c}{Parallel}&\multicolumn{2}{c}{Sequential}\\
  &&&&&&Average&Model-based&$WCX|C18$&$C18|WCX$\\
  \hline
     & Brier Score     & 0.128 & 0.111 && 0.113 & 0.109&0.093  &0.088  &0.092 \\
   & Deviance     & 253.36 & 227.25 && 231.07 & 226.92 &201.00  &188.50  &199.55 \\
    Ridge & c-index   & 0.879 & 0.911 && 0.900 & 0.925 & 0.922 & 0.933&0.922 \\
     & $Q^2$  & 0.304& 0.360 && 0.354 & 0.286 &0.559  &0.519  &0.537  \\
     \hline
          & Brier Score     &0.120  &0.091  &&0.092  &0.087  &0.079  &0.086  &0.083  \\
     & Deviance     & 253.46&198.50  &&204.41  & 190.30 &179.85 &209.63  &192.16 \\
    Lasso& c-index  & 0.877 &0.929  &&0.919  & 0.939 & 0.939 &0.920 &0.939 \\
   & $Q^2$  &0.522  &0.567  && 0.582&0.463  &0.609  &0.673  &0.703  \\
    \hline
    \end{tabular}%
  \label{tab:addlabel}%
\end{table}%

\begin{table}[htbp]
  \centering
  \caption{Breast cancer data}
    \begin{tabular}{cccccccccc}
    \hline
          &    \multicolumn{3}{c}{Re-calibrated single source} && \multicolumn{5}{c}{Combination methods} \\
          \hline
  &&C18&WCX&&Naive&\multicolumn{2}{c}{Parallel}&\multicolumn{2}{c}{Sequential}\\
  &&&&&&Average&Model-based&$WCX|C18$&$C18|WCX$\\
  \hline
     & Brier Score     & 0.126 & 0.107 && 0.113 & 0.109&0.093  &0.088  &0.092 \\
   & Deviance     & 249.69 & 224.47 && 231.07 & 226.92 &201.00  &188.50  &199.55 \\
    Ridge& c-index  & 0.875 & 0.906 && 0.900 & 0.925 & 0.922 & 0.933&0.922 \\
     & $Q^2$  & 0.441& 0.500 && 0.354 & 0.286 &0.559  &0.519  &0.537  \\
     \hline
          & Brier Score     &0.123  &0.093  &&0.092  &0.087  &0.079  &0.086  &0.083  \\
     & Deviance     & 254.12&205.96  &&204.41  & 190.30 &179.85 &209.63  &192.16 \\
    Lasso& c-index  & 0.872 &0.926  &&0.919  & 0.939 & 0.939 &0.920 &0.939 \\
   & $Q^2$  &0.430  &0.549  && 0.582&0.463  &0.609  &0.673  &0.703  \\
    \hline
    \end{tabular}%
  \label{tab:addlabel}%
\end{table}%

\subsubsection{DILGOM data}
In the DILGOM data, each of the considered sources of predictors (transcriptomics and metabolomics) presents a considerably different performance. In terms of discrimination, the metabolome itself presents a c-index around 0.81 while the discriminatory ability of the transcriptomics is notably lower with c-index around 0.70. In the same line, we observed that the metabolomics predictors also outperform transcriptomics in terms of calibration. Note that the $ Q^{2}$ of metabolomics is more than twice larger than the $Q^{2}$ for transcriptomics. These differences are observed with both studied methods, ridge and lasso regression.

\begin{table}[htbp]
  \centering
  \caption{DILGOM data}
    \begin{tabular}{cccccccccc}
    \hline
          &       & \multicolumn{2}{c}{Single source} && \multicolumn{5}{c}{Combination methods} \\
          \hline
  &&Transc&Metab&&Naive&\multicolumn{2}{c}{Parallel}&\multicolumn{2}{c}{Sequential}\\
  &&&&&&Average&Model-based&$Metab|Transc$&$Transc|Metab$\\
  \hline
     & Brier Score     & 0.142 & 0.116 && 0.134 & 0.120 &0.114  &0.113  &0.114  \\
   & Deviance     & 363.66 & 306.95 && 340.64 & 314.47 &300.04  &298.32  &304.20  \\
    Ridge& c-index   & 0.716 & 0.811 && 0.790 & 0.837 & 0.827 & 0.829&0.815  \\
     & $Q^2$  & 0.057 & 0.256 && 0.090 & 0.102 & 0.285 &0.284  & 0.286 \\
     \hline
          & Brier Score     & 0.146 & 0.121 && 0.132 & 0.123 &0.121  &0.123  &0.124  \\
     & Deviance     & 371.95 & 311.09 && 337.77 & 318.37 &311.09 &319.62  &315.15 \\
    Lasso& c-index   & 0.682 & 0.806 && 0.768 & 0.808 &0.806  & 0.808 &0.803 \\
   & $Q^2$  & 0.108 & 0.285 && 0.201 & 0.128 & 0.285 & 0.359 &0.295  \\
    \hline
    \end{tabular}%
  \label{tab:addlabel}%
\end{table}%

The differences between metabolomics and transcriptomics remain approximately the same when focusing on re-calibrated single-omic predictions (Table 4). In the DILGOM setting, re-calibration only improves the results obtained for transcriptomics  based on ridge regression. The performance of the re-calibrated ridge-based model for metabolomics is slightly worse than the crude ones. As we observed in the breast cancer settings, the re-calibration of single-source lasso-based predictions is not beneficial.

The performance of the na\"{i}ve method is clearly unsatisfactory, specially in terms of calibration for the ridge specification. Focusing on $Q^{2}$ as summary measure, the behavior of the na\"{i}ve combination ($Q^{2}=0.090$) is far from the performance of the best single-source model, based on metabolome, with $Q^{2}=0.256$. Similar results were found with regard to the Brier score and the deviance. Regarding calibration, the performance of the na\"{i}ve combination approach, even if sub-optimal, is better when it relies on a lasso specification. For example, the na\"{i}ve combination based on lasso regression presents $Q^{2}=0.201$, while for the metabolome-based model, we found $Q^{2}=0.285$. In terms of c-index, both lasso and ridge na\"{i}ve combinations behave similarly.

As in the breast cancer setting, we observe a better performance of the averaging parallel combination in terms of discrimination, while the model-based parallel combination outperforms averaging in terms of calibration. The sequential approaches behave similar to the model-based parallel combination, with a slight outperformance of the sequential approaches based on using transcriptomics first.

Specifically, even if the differences are slight, it seems that the strategy of first fitting a model based on transcriptomics and adding metabolome in a second step is the preferable one, for both ridge and lasso specifications. Moreover, lasso seems to provide better estimates of calibration than ridge ($Q^{2}=0.359$ for lasso versus $Q^{2}=0.284$ for ridge), while the ridge specification provides larger discrimination ability (c-index=0.829 for ridge versus c-index=0.808 for lasso). Nevertheless, for the ridge specification the maximum c-index is reached by the average parallel combination (c-index=0.837). For lasso regression, both average parallel and the sequential combination $Metab|Transc$ provide the same maximum c-index=0.808.

However, the benefit of using a combination of transcriptomics and metabolomics instead of using metabolomics alone in order to classify individuals as obese or not is arguable from the discrimination point of view, since the differences in c-index are small. The improvement in calibration is also not clear, only in terms of $Q^{2}$ the combination of both sources seems to outperform the prediction based on metabolites only.
\begin{table}[htbp]
  \centering
  \caption{DILGOM data}
    \begin{tabular}{cccccccccc}
    \hline
          &        \multicolumn{3}{c}{Re-calibrated single source} && \multicolumn{5}{c}{Combination methods} \\
          \hline
  &&Transc&Metab&&Naive&\multicolumn{2}{c}{Parallel}&\multicolumn{2}{c}{Sequential}\\
  &&&&&&Average&Model-based&$Metab|Transc$&$Transc|Metab$\\
  \hline
     & Brier Score     &0.142  &0.117 && 0.134 & 0.120 &0.114  &0.113  &0.114  \\
   & Deviance     &361.63  &313.08  && 340.64 & 314.47 &300.04  &298.32  &304.20  \\
    Ridge& c-index   &0.711 & 0.805 && 0.790 & 0.837 & 0.827 & 0.829&0.815  \\
     & $Q^2$  &0.111  &0.247  && 0.090 & 0.102 & 0.285 &0.284  & 0.286 \\
     \hline
          & Brier Score     &0.147  &0.122  && 0.132 & 0.123 &0.121  &0.123  &0.124  \\
     & Deviance     &374.38  &314.38 && 337.77 & 318.37 &311.09 &319.62  &315.15 \\
    Lasso& c-index   &0.676  &0.801  && 0.768 & 0.808 &0.806  & 0.808 &0.803 \\
   & $Q^2$  &0.084& 0.245 && 0.201 & 0.128 & 0.285 & 0.359 &0.295  \\
    \hline
    \end{tabular}%
  \label{tab:addlabel}%
\end{table}%

\section{Summary and discussion}
In this work, we have addressed the problem of integrating several omic sets in the context of prediction of binary outcomes. Several methods for combination of single-source predictions have been presented, all relying on regularized regression.

First of all, we have considered two `parallel' combination approaches, in which a new vector of predictions is obtained as a weighted sum of single-source predictions of the outcome of interest. The specific weight may be fixed beforehand (for example averaging the single-source predictions) or estimated (model-based parallel combination). For the latter, we considered a logistic regression model using the individual predictions as covariates. Given that the single-source predictions are fitted themselves, the model-based combination requires to be embedded in the cross-validatory setting in order to obtain unbiased final combined predictions.

As an alternative to parallel approaches, we have considered a `sequential' combination method consisting of choosing beforehand one of the omic sources as `primary'. Namely, we propose to introduce the vector of individual predictions based on the `primary' source as an extra covariate with fixed weight when fitting a prediction model based on the `secondary' source of omic predictors. In the context of logistic regularized regression models, this is implemented by considering the vector of predictions based on the `primary' source as an offset term. As in the model-based parallel combination, the use as covariate of a vector of predictions (which are fitted themselves) requires an extra layer of cross-validation to embed the uncertainty of calibrating the first source of predictors in the procedure.

We have applied the studied combination methods to two omic applications. Firstly, we have revisited the problem of the combination of different fractionations in proteomic spectometry for breast cancer diagnosis. Secondly, we have evaluated the possible combination of transcriptomics and metabolomics for the prediction of obesity.

Our results show that better predictions can be obtained by combining predictions based on different omic sources, outperforming single-omic predictions. This seems to be the case for the first of our applications, as combining two proteomic fractions benefits breast cancer classification, from both discrimination and calibration points of view. With respect to the DILGOM study, our results are less conclusive. The sequential approach suggests that transcriptomics are of little use for improving the predictive performance of a ridge or lasso model based on metabolomics only. The reverse is not true, as enriching a transcriptomic-based predictor with metabolomics measures leads to more accurate predictions.

The preferable method to conduct such combination seems to depend on the aspect we focused on for the evaluation of the resulting models. In this work, we have evaluated the resulting predictions in terms of discrimination and calibration. For improving discrimination, measured through c-index, averaging single predictions seems to be enough, and in fact, this simple method provides slightly better results than the more sophisticated  model-based parallel approach. In terms of calibration, both parallel and sequential model-based approaches present better and comparable (between them) performance than more simple approaches, as averaging of individual predictions. As expected, the na\"{i}ve approach, consisting of stacking the omic sources ignoring their different origin, is highly misleading.

We have focused on combination approaches for the integration of different sets of omic predictors. An alternative route, still based on regularized regression,  may be to consider different penalizations for each different omic dataset \citep{Meier, Wiel}. A systematic comparison of these methods with combination approaches is left as future research. Also, the analysis in this context of alternative model building techniques which (to some extent) rely on the idea of combination of simple classifiers such as random forests \citep{Breiman} and boosting \cite{Buhlmann, Tutz} is left as interesting line of future research. Also, it would be interesting to consider modifications of the currently used  model-based parallel combination. For example, positively-restricted regression coefficients  \citep{Alexia} or non-linear combinations of single-source predictions could be considered.

Finally, we would like to highlight that we have focused on evaluating the predictive ability of combinations of different omic-based predictions. Formally testing the added predictive value of a given omic dataset on top of an established one falls beyond the scope of this work. Recently, a test for added value based on the sequential approach presented on Section 2.3. has been proposed for continuous outcomes \citep{Mar2}. Its extension to binary outcome is left as a promising line of future research.

\section*{Acknowledgements}
Work supported by funding from the European Community’s Seventh Framework Programme FP7/2011: Marie Curie Initial Training Network MEDIASRES with the Grant Agreement Number 290025 and by funding from the European Union’s Seventh Framework Programme FP7/Health/F5/2012: MIMOmics under the Grant Agreement Number 305280. We thank Sigrid Jusélius Foundation and Yrjö Jahnsson Foundation for providing the DILGOM expression data and Yrjö Jahnsson Foundation for providing  DILGOM NMR metabolomics.

\end{document}